\documentclass[10pt,conference]{IEEEtran}
\usepackage{graphicx} 

\usepackage{orcidlink}
\usepackage{cite}
\usepackage{underscore}
\usepackage[T1]{fontenc}
\usepackage[utf8]{inputenc}
\usepackage{multirow}

\title{Leveraging Large Language Models for Use Case Model Generation from Software Requirements}
\author{
\IEEEauthorblockN{Tobias Eisenreich\orcidlink{0009-0004-7168-251X}\IEEEauthorrefmark{1}, Nicholas Friedlaender\orcidlink{0009-0009-3485-9861}\IEEEauthorrefmark{1}, Stefan Wagner\orcidlink{0000-0002-5256-8429}}
\IEEEauthorblockA{School of Computation, Information and Technology\\Technical University of Munich\\\IEEEauthorrefmark{1}These authors contributed equally to this work.}
}

\usepackage{booktabs} 
\usepackage{enumitem}
\usepackage{calc}
\usepackage{csquotes}
\usepackage{tabularx}

\usepackage{pgfplots}
\pgfplotsset{compat=1.8}
\usepgfplotslibrary{statistics}

\newlist{questions}{enumerate}{2}
\setlist[questions]{label=\textbf{RQ\arabic*:},ref=RQ\arabic*,labelindent=0cm,leftmargin=*}

\begin{document}

\maketitle

\renewcommand{\thefootnote}{}
\footnotetext{© 2025 IEEE. Personal use of this material is permitted. Permission from IEEE must be obtained for all other uses, in any current or future media, including reprinting/republishing this material for advertising or promotional purposes, creating new collective works, for resale or redistribution to servers or lists, or reuse of any copyrighted component of this work in other works.}
\renewcommand{\thefootnote}{\arabic{footnote}}

\begin{abstract}
Use case modeling employs user-centered scenarios to outline system requirements. These help to achieve consensus among relevant stakeholders. Because the manual creation of use case models is demanding and time-consuming, it is often skipped in practice. This study explores the potential of Large Language Models (LLMs) to assist in this tedious process. The proposed method integrates an open-weight LLM to systematically extract actors and use cases from software requirements with advanced prompt engineering techniques. The method is evaluated using an exploratory study conducted with five professional software engineers, which compares traditional manual modeling to the proposed LLM-based approach. The results show a substantial acceleration, reducing the modeling time by 60\%. At the same time, the model quality remains on par. Besides improving the modeling efficiency, the participants indicated that the method provided valuable guidance in the process.


\end{abstract}

\section{Introduction}
Use case modeling serves as a comprehensive toolkit for understanding, documenting, and analyzing the functional view of a system, taking into account the interactions between the system and actors \cite{bittner2003use}. It provides a structured approach to capturing requirements in functional scenarios that depict user-centric goals. This improves communication between stakeholders and developers, providing a common understanding of the system's functional requirements independent of the lower-level implementation.

Use case modeling is a process of requirements analysis in software engineering. It is usually employed in the early phases of system development as part of the requirements capturing and specifications phase. The primary steps in the conventional use case generation process are (1) the identification of the actors, (2) the definition of the use cases, and (3) the establishment of the relationships between them. Actors include users, hardware, and other systems that exist outside the system boundaries and interact with it. The potential benefits of use case models include stakeholder alignment, as they can visualize and understand the functionality of the planned system more easily. Additionally, it allows software maintainers to comprehend the enabled functionality of the system. Use case models provide the basis for designing and testing applications by clearly defining the system's scope. As of now, a detailed and accurate development of use case models often demands a considerable investment of time and effort, due to the domain knowledge and modeling skills required.

The emergence of LLMs represents a promising opportunity for improving the efficiency and accessibility of use case modeling. The LLMs' ability to process complex textual information in natural language makes them particularly well-suited for tasks requiring generating and refining structured output directly from textual input. Trained on diverse and large datasets, LLMs show advanced capabilities in performing different tasks, ranging from coherent and contextually relevant text generation to solving complex problem-solving queries and generating executable code \cite{Li2021}.

Eisenreich et al.~\cite{Eisenreich2024,eisenreich2025requirements} outlined a novel approach to employ LLMs in generating software architectures semi-automatically from requirements. Their vision paper focuses on the potential of LLMs to bridge the gap between requirements engineering and software architecture. The generation of domain and use case models is emphasized in an intermediate step to formalize requirements. Building on this work, this study presents a novel method that harnesses the capabilities of LLMs to assist in creating use case models from textual requirements. It incorporates advanced prompt engineering techniques to transform textual software requirements into use case models.

Related research investigated the feasibility of leveraging LLMs for UML (Unified Modeling Language) model generation \cite{wang2024llms,camara2023assessment,ferrari2024model} and indicates promising results. This research further builds a method, i.e., a process and tool, around the use case model generation task. Implementing and evaluating the proposed method provides an in-depth assessment involving software engineers to address the following research questions:
\begin{questions}
\item How much can the proposed method accelerate the development of use case models?
\item How do developers perceive the LLM-based assistance when creating use case models?
\end{questions}

The remainder of this paper is structured as follows:
In \autoref{sec:bgr}, we discuss the background that supports this study and the related work.
In \autoref{sec:methods}, the supporting tool is explained in detail, along with the methodology for conducting the evaluation.
The results are presented in \autoref{sec:results} and further discussed in \autoref{sec:discussion}.
Finally, we discuss the threats to validity in \autoref{sec:threats} and conclude the paper in \autoref{sec:conclusion}.

\section{Background and Related Work}\label{sec:bgr}
This section explains the background on use case models and discusses related work on using LLMs in UML modeling.
\subsection{Use Case Model}
Use case modeling is a technique for describing functional requirements from a user's perspective \cite{hofmann2001requirements}. It connects stakeholders and developers, fostering alignment and understanding by exhibiting how a system interacts with its actors. The generation of the use case model is typically initiated during the requirements analysis phase and is primarily conducted under the supervision of the Product Owner (PO).

A use case model typically has two components: (1) a use case diagram and (2) use case descriptions \cite{usecasediagram}. A use case diagram is usually modeled in UML and consists of four components: actor(s), use cases, system boundary, and associations \cite{bittner2003use}.
A use case description complements the use case diagram by detailing the sequence of actions that the system performs to attain a certain use case. \cite{bittner2003use}.

\subsection{Related Work}

As the LLM-based generation of use cases is a relatively novel area of study, the scope of this review includes related research on the general generation of UML diagrams from textual requirements. This broader focus provides insights into how LLMs interpret textual inputs for modeling tasks and their capabilities in entity extraction.

Wang et al.\cite{wang2024llms} investigated how LLMs support software engineering students in understanding and creating UML diagrams. They show that LLMs can aid in generating class, use case, and activity diagrams, but the effectiveness depends heavily on prompt quality and the LLMs' domain-specific knowledge. The researchers highlighted limitations of current LLMs, such as inherent randomness during interactions. They propose prompt engineering to better control and refine LLM outputs. Our work builds upon this by applying prompt templates that are enhanced with concrete requirements.

Li et al. \cite{li2024llm} examined the potential of LLMs guided by chain-of-thought (CoT) prompting for automating the translation of user stories into UML class diagrams. They found that LLMs are effective in identifying classes from user stories, particularly when combined with well-crafted prompts. Observed limitations include the LLM sometimes failing to recognize important domain entities, indicating a lack of deep semantic understanding.

Ferrari et al. \cite{ferrari2024model} explored LLMs for generating UML sequence diagrams directly from textual requirements. The findings emphasize the importance of interactive prompting and feedback loops to enhance the accuracy and relevance of the generated diagrams. This interactive approach is highlighted as essential for producing high-quality results. We adopt this approach in our work.

Similarly, Cámara et al. \cite{camara2023assessment} critically evaluated ChatGPT's performance in UML modeling tasks. They acknowledged strengths like rapid generation of initial drafts and identified limitations, such as difficulties when handling complex or ambiguous requirements.

The importance of prompt engineering to enhance LLM performance is supported by White et al. \cite{white2023promptpatterns}. They presented a collection of prompt patterns tailored for software engineering tasks. The study's findings demonstrate that using well-crafted prompts drastically enhances the accuracy and usability of LLM outputs.

The presented studies demonstrate the potential of LLMs in UML modeling. However, existing work does not outline a process integrating LLMs and prompt engineering for use case model generation. No studies have quantitatively assessed the extent to which LLM-based approaches can accelerate the modeling process. There is also a lack of in-depth qualitative analysis capturing software engineers' experiences and perceptions when using LLMs for use case modeling.

\section{Methodology}\label{sec:methods}
This section first explains the high-level research design. We then discuss the implementation of the supportive tooling. Finally, we detail the experiment procedure as well as the data analysis.
\subsection{Research Design and Strategy}
A mixed methods approach was chosen to evaluate the proposed methods' effectiveness and user experience for generating LLM-based use case models. This approach combines quantitative and qualitative research studies to comprehensively understand how software engineers perceive the LLM-based modeling approach, supporting this with an experiment and interviews.

To answer \textit{RQ1}, we collected numerical data that supports the observed time differences between the two modeling approaches, as well as data on the model quality, which compares accuracy between manual and AI-generated models. The quantitative component of the study involved measuring the time it took participants to develop a use case model, both with and without the proposed method. This was based on predefined requirements texts of comparable length. To improve accuracy, the requirement texts were cross-validated by substituting them among the participants. The within-subject experimental design minimizes inter-participant variability, as each participant completed the exercise in both conditions, ensuring direct comparability.

The qualitative component consisted of a semi-structured interview to explore \textit{RQ2}. The semi-structured interview was chosen to explore topics in more depth by asking follow-up questions. This flexibility allowed the interviewer to adapt to the conversation and examine interesting or unexpected responses more closely, providing more detailed information.

Participants were recruited contingent on recommendations from professional networks. The participant pool had an average of seven years of relevant professional software development experience, ensuring they had the necessary knowledge, skills, and expertise for the study.

The mixed methods approach improves the understanding of the experiment results and interview findings. Employing different data types strengthens the validation process and reinforces the overall argument \cite{blank2013sage}. This approach ensures that both \textit{RQ1} and \textit{RQ2} are addressed comprehensively. While the modeling time may be used to compare performance, qualitative feedback can explain the underlying reasons for these differences, improving the interpretation and application of results \cite{johnson2004mixed}.

\subsection{Tool}
For this study, a method, i.e., a process and tool, was developed for the generation of use case models from requirements text. The tool stands at the core of this study's experiments and interviews, designed around the use case model generation process.

The developed tool encompasses the unmodified open-weight LLM Llama 3.1 70B~\cite{grattafiori2024llama}, for entity extraction and for generating the use case model in PlantUML\footnote{\url{https://plantuml.com}}. To improve the LLMs' output, different prompt engineering patterns were employed: Role prompting \cite{xu2023expertprompting} is implemented by instructing the LLM to act as an expert in software engineering. Knowledge Injection \cite{martino2023knowledge} provides the LLM with comprehensive knowledge in PlantUML syntax, and Negative Prompting \cite{miyake2023negativeprompt} helps to avoid common errors that can occur during the generation of use case models.

After initial testing, we opted for a semi-automated modeling workflow, as an automated process frequently resulted in the model hallucinating and inaccurate PlantUML syntax. This decision aligns with the recommendations of Tiwari et al. \cite{tiwari}, who advocate for a semi-automated questionnaire method to empower users with enhanced control over the output.

\begin{figure}[tbp]
\centering
\includegraphics[width=\columnwidth]{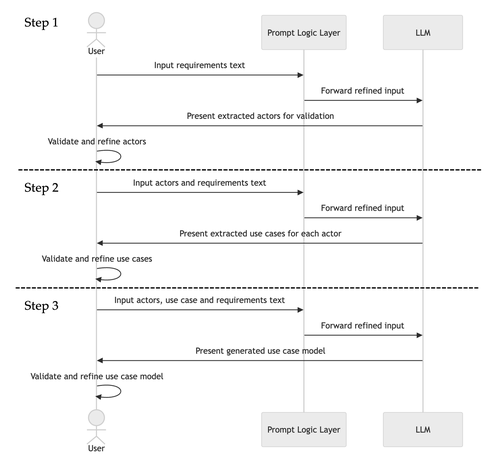}
\caption{Proposed LLM-based use case modeling workflow}
\label{fig:method}
\end{figure}

\autoref{fig:method} shows the workflow of the developed tool:
In the first step, all actors from the requirements text are identified. In the next step, only use cases that have a relationship with one or more of these actors are extracted. Finally, based on the intermediate results, the use case model is generated. Before each step, users have the opportunity to validate and refine the results from the previous step.

 The repository containing the implemented method, including the prompts and requirements texts, is available on Zenodo~\cite{friedlaender2025code}.

\subsection{Experiment Procedure}
The study consisted of four distinct phases, which took approximately 50 minutes to conduct. All participants completed the tasks in the following order:

\textit{Step 1 - Introduction:}
To ensure familiarity with use case modeling, participants received an overview of the research topic and an example use case modeling exercise.

\textit{Step 2 - Baseline exercise:}
Participants were assigned one of the two requirement texts and manually created a use case diagram, plus detailed descriptions for two use cases from the diagram.

\textit{Step 3 - LLM-based exercise:}
Subsequently, participants were given the remaining requirement text and used the proposed method to develop the use case model.

\textit{Step 4 - Semi-structured interviews:}
Finally, after completing the prior tasks, participants were interviewed, investigating the following areas: (1) impressions and usability, (2) usefulness, (3) quality and reliability of outputs, (4) challenges, and (5) suggestions for improvement.

\subsection{Data Collection and Analysis}

The quantitative data collected is the amount of time required by the participant to complete the modeling in steps 2 and 3. For both modeling steps, the participants had no time limit and were instructed to model until they were satisfied with their results. The time spent reading through the textual requirements for the first time is not included in the measured times. Omitting this variable from the calculations helped to reduce the impact of external factors, such as individual reading speed. The qualitative data from the interviews were audio-recorded and later transcribed for subsequent analysis.

Even with the limited sample size of five observations, a paired t-test can successfully find a significant difference between the two modeling conditions if the difference is high enough. The aim of this statistical analysis was to provide preliminary empirical evidence on the effects of the tool.

To further support the statistical analysis of RQ1, we evaluated the quality of the generated use case models against each participant's manual approach. This evaluation involved comparing the models to a manually created ground truth model, based on semantic alignment. The evaluated data is available in the supplementary material \cite{friedlaender2025code}. The quality score is based on the core model elements -- actors and use cases -- which form the structural foundation of a use case model. It does not include relationships, because their interpretation is more context-dependent. The detailed descriptions of the use cases are hard to measure and were not included either.

The qualitative data were structured with a thematic analysis, which identified recurring patterns, insights, and suggestions. The interviews were transcribed and coded using a systematic marking process to generate key themes as proposed by Braun et. al.~\cite{Braun2006}.

\section{Results}\label{sec:results}

This section lists the results of the quantitative and qualitative analysis.

\subsection{Quantitative Analysis}
\begin{table}[tbp]
\centering
\begin{tabular}{ccccc}
\toprule
\textbf{Number} & \textbf{Requirements} & \textbf{Exercise} & \textbf{Time (min)} \\
\midrule
P1 & B & Manual    & 14.2 \\
P2 & A & Manual    & 19.1 \\
P3 & B & Manual    & 11.02 \\
P4 & B & Manual    & 16.9 \\
P5 & A & Manual    & 26.3 \\
P1 & A & LLM-based & 4.8 \\
P2 & B & LLM-based & 6.25 \\
P3 & A & LLM-based & 4.2 \\
P4 & A & LLM-based & 9.6 \\
P5 & B & LLM-based & 10.4 \\

\bottomrule
\end{tabular}
\vspace{0.5em}
\caption{Unprocessed observed modeling data}
\label{tab:time}
\end{table}

The within-subject experiment involved five participants, each tasked with generating a use case model for two similar requirements texts, one manually and one utilizing the proposed LLM-based method. The modeling times were recorded for both methods to evaluate efficiency gains. The results show that the LLM-based approach significantly accelerates the use case modeling process. On average, participants completed the task roughly 60\% faster using the LLM-based method.

A paired t-test was conducted to determine whether the observed differences in modeling times between the manual and LLM-based approaches were statistically significant. The prerequisite that the data follows a normal distribution was evaluated using the Shapiro-Wilk test\cite{shaphiro1965analysis}. The paired t-test  yields a t-statistic of $6.05$ with a p-value of $0.0037$, indicating that the time reduction associated with the LLM-based approach is statistically significant at the $\alpha = 0.01$ level.

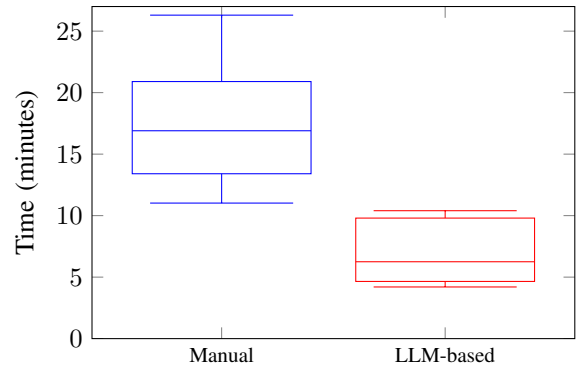
\begin{figure}[tbp]
\centering
\begin{tikzpicture}
  \begin{axis}[
    boxplot/draw direction=y,
    xtick={1,2},
    xticklabels={Manual, LLM-based},
    x tick label style={font=\footnotesize, rotate=0, anchor=north},
    ylabel={Time (minutes)},
    height=6cm,
    width=8cm,
    ymin=0,
    ymax=27,
    ytick={0,5,10,15,20,25}
    ]
    \addplot+[
      boxplot prepared={
        median=16.9,
        upper quartile=20.9,
        lower quartile=13.405,
        upper whisker=26.3,
        lower whisker=11.02
      },
    ] coordinates {};

    \addplot+[
      boxplot prepared={
        median=6.25,
        upper quartile=9.8,
        lower quartile=4.65,
        upper whisker=10.4,
        lower whisker=4.2
      },
    ] coordinates {};
  \end{axis}
\end{tikzpicture}

\vspace{0.5em}

\caption{Box plot of the modeling time for the Manual and LLM-based approach}
\label{tab:boxplot}
\end{figure}

Despite the small sample size and due to the large effect observed, the statistic analysis concludes that the proposed method significantly improves efficiency in use case modeling. Due to the lack of counterbalancing (see \autoref{sec:threats}), the results should be regarded with caution. Detailed time measurements and descriptive statistics are summarized in \autoref{tab:time} and \autoref{tab:boxplot}, respectively. Additionally, \autoref{tab:usecases_metrics} exhibits high precision, indicating most use cases generated by the model were valid, while a higher recall compared to the manual approach suggests that more elements from the ground truth were identified, further supporting the quality of the results. Since all actors were correctly identified in both approaches, no separate table was provided for these calculations.

\begin{figure}[tbp]
\centering
\includegraphics[width=\columnwidth]{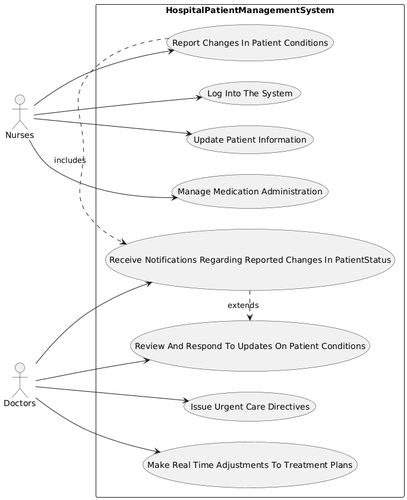}
\caption{Use Case Diagram – LLM-based Method (Participant 2)}
\label{fig:result}
\end{figure}

\subsection{Qualitative Analysis}
A thematic analysis of interview data \cite{Braun2006} complements the quantitative analysis. Four key themes provide a deeper understanding of participants' experiences and perceptions of the method: (1) Perceived Usefulness, (2) Quality and Reliability of Outputs, (3) Perceived Concerns, and (4) Proposed Areas for Improvement. A summary of the key insights of each theme is presented in \autoref{tab:insights}.

\begin{table*}[tbp]
\centering
\small
\begin{tabularx}{\textwidth}{lX}%
\toprule
\textbf{Theme} & \textbf{Key Insights} \\
\midrule
Perceived Usefulness &
Simplifies the modeling process and accelerates development. Lightens workload, making use case modeling more accessible to non-experts. \\
\addlinespace
Quality and Reliability &
Outputs align closely with expectations. Provides the user with a feeling of control. \\
\addlinespace
Perceived Concerns &
Lacks iterative dialogue. Risk of over-reliance on outputs. Inconsistencies across iterations and limited domain adaptability. \\
\addlinespace
Proposed Improvements &
Suggestions include an interpretability feature, output caching, and a visual UML editor. \\
\bottomrule
\end{tabularx}
\vspace{0.5em}
\caption{Themes and key insights from the qualitative analysis}
\label{tab:insights}
\end{table*}

\renewcommand{\arraystretch}{1.3}
\begin{table}[tbp]
\centering
\begin{tabular}{l@{\hskip 5pt}ccc|ccc}
\toprule
\multirow{2}{*}{\textbf{Participant}}
& \multicolumn{3}{c}{\textbf{Manual}}
& \multicolumn{3}{c}{\textbf{LLM}} \\
& \textbf{Precision} & \textbf{Recall} & \textbf{F1}
& \textbf{Precision} & \textbf{Recall} & \textbf{F1} \\
\midrule
P1 & 1.00 & 0.83 & 0.91 & 0.88 & 0.78 & 0.82 \\
P2 & 0.88 & 0.78 & 0.82 & 0.75 & 1.00 & 0.86 \\
P3 & 0.83 & 0.83 & 0.83 & 0.67 & 1.00 & 0.80 \\
P4 & 0.67 & 0.67 & 0.67 & 0.82 & 1.00 & 0.90 \\
P5 & 0.75 & 0.67 & 0.71 & 0.86 & 1.00 & 0.92 \\
\midrule
\textbf{Avg}
& \textbf{0.83} & 0.76 & 0.79
& 0.79 & \textbf{0.96} & \textbf{0.86} \\
\bottomrule
\end{tabular}
\vspace{1em}
\caption{Precision, Recall, and F1 Score for Manual vs. LLM Approach — \textbf{Use Cases}}
\label{tab:usecases_metrics}
\end{table}

\subsubsection*{Theme 1: Perceived Usefulness}\phantom{.}\\
A key theme that emerged from the interviews is the perceived usefulness of the proposed LLM-based method. All participants were optimistic about the method's usefulness and emphasized its ability to reduce the workload of software engineers and significantly speed up the modeling process.

The participants emphasized several benefits of the workflow: The method helps to refine intermediate results by eliminating irrelevant or hallucinated data generated by the LLM in earlier phases. It also allows software engineers to confirm or revise the model's interpretation of the textual requirements. Rather than starting from scratch, a user can build on the generated model, refining and adapting it as required.

The participants underlined the perceived benefits of the LLM-based use case model generation. They noted that the proposed method significantly accelerates the creation of use case models compared to manual approaches. The tool was perceived as particularly valuable for teams with limited resources who want to improve their development practices. Participant 4 said that \enquote{Despite the usefulness of use case diagrams, they are rarely created in practice due to a lack of expertise or time.} Similarly, participant 5 commented: \enquote{The tool is particularly useful for application lifecycle management, enabling non-experts in requirements engineering to design requirements or identify users and their interactions with the system.}

\subsubsection*{Theme 2: Quality and Reliability of Outputs}\phantom{.}\\
The second important theme raised in the interviews was the quality and reliability of the results generated using the proposed method. The participants rated the method's performance positively, especially regarding how well the generated results aligned with the requirement text. Overall, they were satisfied with the system's ability to produce reliable results while leaving room for user input and control.

The participants praised the method for producing results similar to those they might have created manually. It was characterized by a high level of alignment with the content and intent of the requirements text and provided users with control throughout the modeling process. Overall, participants were satisfied with the performance, which provided reliable outputs while giving ample opportunities for user intervention. Participant 1 described the results as \enquote{meaningful}, and participant 3 commented: \enquote{I was satisfied with the generated use cases and would have modeled them similarly myself}.

\subsubsection*{Theme 3: Perceived Concerns}\phantom{.}\\
This theme reflects the challenges and limitations participants identified in applying the proposed LLM-based method. These concerns include both practical workflow issues and LLMs' inherent limitations, particularly in real-world software development practice.

Participants noted that the method lacks reasoning capabilities to check whether the generated use cases match the requirements text. Unlike traditional requirements engineering, which typically involves iterative and collaborative processes, the tool does not support dialogue between stakeholders to uncover and refine the underlying goals. Participant 5 explained: \enquote{A requirements engineer usually questions assumptions, asks why, and explores alternative solutions. This iterative process is missing from the workflow.}

Several participants also expressed concerns that users may be less inclined to critically engage with the results. They pointed out that the tool assumes precisely defined requirements, a condition rarely met in real-world projects. Participant 2 noted: \enquote{The scalability of the current version needs to be validated with larger volumes of requirements text.}

Participants identified two technical shortcomings: First, the non-de\-ter\-mi\-nism of LLMs causes inconsistency across iterations. Second, the results were often not detailed enough in the initial generation of the use case descriptions and needed manual expansion.

\subsubsection*{Theme 4: Proposed Areas for Improvement}\phantom{.}\\
The final theme identified in the interviews highlights suggestions to improve workflow, quality of outcomes, and user engagement. Participants made several specific propositions to improve both the user experience and the overall effectiveness of the LLM-based method.

Most participants recommend adding a feature that allows users to trace use cases back to specific sections of the requirements text. This would support users in the manual validation and help them to understand how the results were derived. Participants also emphasised the importance of transparency and coverage in the modelling process. The tool should generate use cases explicitly mentioned in the requirements and highlight relevant use cases that may have been omitted, along with explanations for their possible exclusion. In addition, the tool should reflect business guidelines and go beyond basic functionality to deliver greater strategic value. Participant 4 suggested: \enquote{A valuable next step would be to generate user stories from use cases.}

   dv                g
   ugozfFinally, creating persona descriptions was recommended to make the modelling process more engaging and refine requirements more effectively. Participant 5 commented: \enquote{Incorporating methods like design thinking with AI-generated persona descriptions or role-playing could encourage active user participation and produce more high-quality results}.

\section{Discussion}\label{sec:discussion}
The quantitative data, answering \textit{RQ1}, shows that the proposed method accelerates the use case modeling time for the study participants significantly. This is consistent with the findings from the interview data, where participants stated that the method reduced the workload. The modeling time is reduced by approximately 60\% on average. At the same time, the model quality remains comparable: The comparison to manually created models shows similar precision, recall, and F1 score. Furthermore, the participants confirm that the generated models have good quality. This can be predominantly ascribed to the incorporation of intermediate steps in the modeling process, which enabled the participants to exercise a high degree of control over the outcome: A close observation of the participants' workflow during the experiment revealed a recurrent tendency to make adjustments to the intermediate or final output of the model: deleting unnecessary actors, editing use case titles, or refining the final output. Our findings support existing literature on the effectiveness of LLMs in use case model development and give an estimate of how much the development times are reduced.

The insights collected during the interviews, answering \textit{RQ2}, indicate that developers are receptive to receiving assistance during the use case modeling process. Participants acknowledged that developing use case models is a fundamental process of requirements engineering. However, it was also noted that this process does not currently receive sufficient attention due to time constraints and go-to-market pressures, resulting in inadequately documented projects. Participants regarded the tool as a potential solution to this issue, as it would alleviate the time burden on developers when creating use case models. The acceleration can lead to more efficient development workflows and improved stakeholder understanding and engagement.
However, participants also noted that in practice, requirements are often developed and refined iteratively over time. Consequently, use case models are typically not created in a single session but evolve alongside the requirements. The participants raised the issue that this evolution is not supported by the developed method. While the participants had a positive attitude toward the tool support, they found a few areas for further improvement of both the method and the tooling.

\section{Threats to validity}\label{sec:threats}
This study acknowledges several limitations that may affect the validity of its findings. First, the limited sample size precludes the possibility of generalizing the findings from the within-subject experiments to the broader community of software engineers~\cite{tipton2017implications}. Future research should use a larger and more varied participant pool to enhance the  robustness and representativeness of the findings.

The complexity and nature of the requirements texts employed in the study constitute an additional limitation. The requirement texts were deliberately crafted to be similar in complexity and length to other UML modeling studies \cite{wang2024llms}. While this design choice enables comparability with related work, it does not fully reflect the challenges encountered in professional domains, where requirements are often significantly longer and more complex. Since the performance of LLMs across different input lengths is not well understood \cite{levy-etal-2024-task}, these results cannot be generalized to more extensive requirements.

When answering RQ1, we found a very large time reduction. However, there is a possible practice effect involved~\cite{greenwald1976within}: Each participant first developed the use case model manually, followed by the LLM-based approach. The authors of this study do not assume that the practice effect can account for the whole effect observed; still, it hurts the statistical significance of the time reduction. To mitigate this effect, a further study should implement counterbalancing or avoid using a within-subjects design.

Finally, the study's results considered both the perceived quality of the LLM-based generated models and an objective measurement of the model quality. The limited sample size prevents drawing definitive conclusions about its effectiveness. Furthermore, the ground truth model created for comparison might not be of sufficiently high quality to create a meaningful metric.

\section{Conclusion}\label{sec:conclusion}

In this paper, we have proposed an LLM-based method for use case modeling and examined both its effectiveness in reducing modeling time and software engineers' attitudes about being assisted in this task. The experimental results showed that the proposed method reduces modeling time compared to manual modeling by 60\% on average. An inspection of the resulting models shows that this method yields a comparable model quality compared to manual modeling. Through follow-up interviews with the study participants, we gained deeper insights into the strengths and limitations of the method. The participants agree that the quality of the generated use case models is comparable to those created manually. The proposed method was well received, and the participants recognized the potential to improve communication between business and IT professionals.

The findings show significant potential in increasing the efficiency of software engineering. This would allow engineers to concentrate on more complex tasks while improving project documentation. A future version of the method should address the limitations identified, increasing the tool's adaptability. Overall, this research demonstrates the potential of LLM-based tools for use case modeling and highlights the need for further research into scaling the method to accommodate larger requirement texts, diverse domains, and collaborative team environments.

\bibliographystyle{IEEEtran}
\bibliography{IEEEabrv,references}

\end{document}